\documentclass[aoas,MSNbibl,nameyear,dvips]{arximspdf}
\usepackage{dcolumn}
\usepackage{graphicx}

\doi{10.1214/11-AOAS458}
\volume{5}
\issue{3}
\pubyear{2011}
\firstpage{2003}
\lastpage{2023}

\makeatletter

\newcolumntype{d}[1]{D{.}{.}{#1}}

\newcommand{\bfD}{\mathbf{D}}
\newcommand{\bfV}{\mathbf{V}}
\newcommand{\bfW}{\mathbf{W}}
\newcommand{\bfX}{\mathbf{X}}
\newcommand{\bfZ}{\mathbf{Z}}
\newcommand{\bftheta}{\bolds{\theta}}
\newcommand{\bfvtheta}{\bolds{\vartheta}}
\newcommand{\bfbeta}{\bolds{\beta}}
\newcommand{\Ttil}{\tilde{T}}
\newcommand{\eqref}[1]{(\ref{#1})}

\let\sv@tabnotetext\tabnotetext
  \let\sv@tabnotemark@fmt\tabnotemark@fmt
   \long\def\legend#1{{\let\tabnote@indent\leavevmode\sv@tabnotetext[]{}{#1}}}
\makeatother

\begin{document}
\begin{frontmatter}

\title{Risk prediction for prostate cancer recurrence through
regularized estimation with simultaneous adjustment for nonlinear
clinical effects\thanksref{T1}}
\runtitle{Risk prediction with regularized partly linear AFT model}
\thankstext{T1}{Supported
in part by the National Institutes of Health Grant  R01 CA106826, the PHS
Grant UL1 RR025008 from the Clinical and Translational Science Award program,
National Institutes of Health, National Center for Research Resources, an
Emory University Research Committee grant, and the Department of Defense IDEA
Award PC093328.}

\begin{aug}
\author[A]{\fnms{Qi} \snm{Long}\corref{}\ead[label=e1]{qlong@emory.edu}},
\author[B]{\fnms{Matthias} \snm{Chung}\ead[label=e2]{conrad@mathcs.emory.edu}},
\author[C]{\fnms{Carlos S.} \snm{Moreno}\ead[label=e3]{cmoreno@emory.edu}}\\
and
\author[A]{\fnms{Brent~A.}~\snm{Johnson}\ead[label=e4]{bajohn3@emory.edu}}
\runauthor{Long, Chung, Moreno and Johnson}
\affiliation{Emory University, Texas State University, Emory
University and~Emory~University}
\address[A]{Q. Long\\
B. A. Johnson\\
Department of Biostatistics \\
\quad and Bioinformatics \\
Emory University\\
Atlanta, Georgia 30322\\
USA \\
\printead{e1}\\
\hphantom{E-mail: }\printead*{e4}} 
\address[B]{M. Chung \\
Department of Mathematics \\
Texas State University\\
San Marcos, Texas 78666\\
USA\\
\printead{e2}}
\address[C]{C. S. Moreno \\
Department of Pathology \\
\quad and Laboratory Medicine \\
Emory University\\
Atlanta, Georgia 30322\\
USA\\
\printead{e3}}
\end{aug}

\received{\smonth{3} \syear{2010}}
\revised{\smonth{1} \syear{2011}}

%
\begin{abstract}
In biomedical studies it is of substantial interest to develop risk
prediction scores using high-dimensional data such as gene expression
data for clinical endpoints that are subject to censoring. In the
presence of well-established clinical risk factors, investigators often
prefer a procedure that also adjusts for these clinical variables.
While accelerated failure time (AFT) models are a useful tool for the
analysis of censored outcome data, it assumes that covariate effects on
the logarithm of time-to-event are linear, which is often unrealistic
in practice. We propose to build risk prediction scores through
regularized rank estimation in partly linear AFT models, where
high-dimensional data such as gene expression data are modeled linearly
and important clinical variables are modeled nonlinearly using
penalized regression splines. We show through simulation studies that
our model has better operating characteristics compared to several
existing models. In particular, we show that there is a nonnegligible
effect on prediction as well as feature selection when nonlinear
clinical effects are misspecified as linear. This work is motivated by
a recent prostate cancer study, where investigators collected gene
expression data along with established prognostic clinical variables
and the primary endpoint is time to prostate cancer recurrence. We
analyzed the prostate cancer data and evaluated prediction performance
of several models based on the extended $c$ statistic for censored
data, showing that (1) the relationship between the clinical variable,
prostate specific antigen, and the prostate cancer recurrence is likely
nonlinear, that is, the time to recurrence decreases as PSA increases
and it starts to level off when PSA becomes greater than 11; (2)
correct specification of this nonlinear effect improves performance in
prediction and feature selection; and (3) addition of gene expression
data does not seem to further improve the performance of the resultant
risk prediction scores.\looseness=-1
\end{abstract}

%
\begin{keyword}
\kwd{Accelerated failure time model}
\kwd{feature selection}
\kwd{Lasso}
\kwd{partly linear model}
\kwd{penalized splines}
\kwd{rank estimation}
\kwd{risk prediction}.
\end{keyword}

\end{frontmatter}

\section{Introduction}\label{sec1}
In biomedical research it is of substantial interest to build
prediction scores for risk of a disease using high-dimensional
biomarker data such as gene expression data for clinical endpoints
subject to censoring, for example, time to the development or
recurrence of a disease. This process typically involves a feature
selection step, which identifies important biomarkers that are
predictive of the risk. When some clinical variables have been
established as the risk factors of a disease, it is preferred to use a
feature selection procedure that also accounts for these clinical
variables. Using observed data with censored outcomes, our goal is to
build risk prediction scores using high-dimensional data through
feature selection while simultaneously adjusting for effects of
clinical variables that are potentially nonlinear.

\subsection{A prostate cancer study}\label{sec1.1}
This article is motivated by a prostate cancer study. An important
challenge in prostate cancer research is to develop effective
predictors of future tumor recurrence following surgery in order to
determine whether immediate adjuvant therapy is warranted. Thus,
biomarkers that could predict the likelihood of success for surgical
therapies would be of great clinical significance. In this study, each
patient underwent radical prostatectomy following a diagnosis of
prostate cancer, and their radical prostatectomy specimens were
collected immediately after the surgery and subsequently formalin-fixed
and paraffin-embedded (FFPE). More recently, the investigators isolated
RNA samples from these specimens and performed DASL (cDNA-mediated
Annealing, Selection, extension and Ligation) expression profiling on
these RNA samples using a custom-designed panel of 1,536 probes for 522
prostate cancer relevant genes. The DASL assay is a~novel expression
profiling platform based upon massively multiplexed real-time
polymerase chain reaction applied in a microarray format, and, more
importantly, it allows quantitative analysis of RNA from FFPE samples,
whereas traditional microarrays do not [\citet{Bibikova04}; \citet{Abramovitz08}]. In addition, important clinical variables were also
collected, two of which, prostate specific antigen (PSA) and total
gleason score, are known to be associated with prostate cancer risk and
prognosis and are of particular interest. The primary clinical endpoint
in this study is time to prostate cancer recurrence. The research
questions of interest include the following: (1) identifying important
probes that are predictive of the recurrence of prostate cancer after
adjusting for important clinical variables; (2) constructing and
evaluating risk prediction scores; and (3)~determining whether the
inclusion of the gene expression data improves the prediction
performance. It was also suspected that PSA may have a nonlinear effect
on the clinical endpoint. In this article we will develop and apply a
new statistical model, which allows us to answer these questions.

\subsection{Feature selection and prediction in AFT}\label{sec1.2}
The accelerated failure time (AFT) model is an important tool for the
analysis of censored outcome data [\citet{Cox84};
\citet{Kalbfleisch02}]. Compared to the more popular proportional
hazard (PH) model [\citet{Cox72}], the AFT model is, as suggested
by Sir David Cox [\citet{Reid94}], ``in many ways more appealing
because of its quite\vadjust{\goodbreak} direct physical interpretation,'' especially when
the response variable is not related to survival time. Furthermore,
when prediction is of primary interest, the AFT model is arguably more
attractive, since it models the mean of the log-transformed outcome
variable, whereas the Cox PH model estimates the hazard functions.

Classic AFT models assume that the covariate effects on the logarithm
of the time-to-event are linear, in which case one could use standard
rank-based techniques for estimation and inference [\citet{Tsiatis90}; \citet{Ying93}; \citet{Jin03}]
and perform a lasso-type [\citet{Tibshirani96}]
variable selection [\citet{Johnson08}; \citet{Cai09}]. Regarding existing
variable selection and prediction procedures, there are two
unsatisfying products. First, the linearity assumption may not hold in
real data. For example, \citet{Kattan03} showed that relaxing the
linearity assumption of the Cox PH model improved predictive accuracy
in the setting of predicting prostate cancer recurrence with
low-dimensional data. Second, an unsupervised implementation of the
regularized variable selection procedure can inadvertently remove
clinical variables that are known to be scientifically relevant and can
be measured easily in practice. We will address both concerns in our
extensions of AFT models.

\subsection{Partly linear models}\label{sec1.3}
It has been well established that linear regression models are
insufficient in many applications and it is more desirable to allow for
more general covariate effects. Nonlinear modeling of covariate effects
is less restrictive than the linear modeling approach and thus is less
likely to distort the underlying relationship between an outcome and
covariates. However, new challenges arise when including nonlinear
covariate effects in regression models. In particular, nonparametric
regression methods encounter the so-called ``curse of dimensionality''
problem, that is, the convergence rate of the resulting estimator
decreases as the dimension of the covariates increases [\citet{Stone80}], which is further exacerbated when the dimension of the
covariates is high. The partly linear model of \citet
{Engle86}; \citet{Hardle00}; \citet{Ruppert03} provides a useful compromise to model the
effect of some covariates nonlinearly and the rest linearly.
Specifically, for the $i$th subject, let $T_i$ be a~univariate endpoint
of interest for the $i$th subject, and $\bfZ_i=(Z_i^{(1)},\ldots
,Z_i^{(d)})^{\mathrm{T}}$ ($d\times1$) and $\bfX_i=(X_i^{(1)},\ldots
,X_i^{(q)})^{\mathrm{T}}$ ($q\times1$) denote high-dimensional features of
interest (say, gene expression levels) and established clinical
variables, respectively. Then one partly linear model of interest is
%
\begin{equation}\label{eq:plm}
T_i=\phi(\bfX_i) + \bfvtheta^{\mathrm{T}} \bfZ_i+ \varepsilon_i,
\end{equation}
where $\bfvtheta=(\vartheta_1,\ldots,\vartheta_d)^{\mathrm{T}}$ is a
parameter vector of interest, $\phi$ is an unspecified function, and
the errors ($\varepsilon_i$) are independently and identically
distributed (i.i.d.) and follow an arbitrary distribution function
$F_{\varepsilon}$. Special cases of this model have been used in
varied applications across many disciplines including econometrics,
engineering, biostatistics and epidemiology [\citet{Hardle00}]. In this
article we consider Model \eqref{eq:plm} for $T_i$ subject to
right-censoring, and, hence, the observed data are $\{(\Ttil_i,\delta
_i, \bfZ_i, \bfX_i)\}_{i=1}^n$, where $\Ttil_i=\min(T_i,C_i)$,
$\delta_i=I(T_i\leq C_i)$, and $C_i$ is a random censoring event. We
note that $T_i$ is the log-transformed survival time in survival
analysis, and we refer to Model \eqref{eq:plm} as partly linear AFT models.

In the absence of censoring, the nonparametric function $\phi$ in
Model (\ref{eq:plm}) can be estimated using kernel methods [\citet{Hardle00}, references therein] and smoothing spline methods
[\citet{Engle86}; \citet{Heckman86}]. For partly linear AFT models, one can extend the
basic weighting scheme of \citet{Koul81}, where one treats censoring
like other missing data problems [\citet{Tsiatis06}] and inversely
weights the uncensored observations by the probability of being
uncensored, that is, so-called inverse-probability weighted (IPW)
estimators. A close cousin to the IPW methodology is censoring unbiased
transformations [\citet{Fan96}, Chapter 5 and references therein],
which effectively replaces a censored outcome with a suitable surrogate
before complete-data estimation procedures are applied. Both IPW
kernel-type estimators and censoring unbiased transformations in the
partly linear model have been studied for AFT models [\citet{Liang98}; \citet{Wang02}]. Since both aforementioned approaches make stronger
assumptions than rank estimation of AFT models [\citet{Cai09}], we
focus on extending rank estimation to meet our needs.

We here consider a general penalized loss function for partly linear
AFT models
%
\begin{equation}\label{eq:pen.loss}
\min_{\bfvtheta,\phi\in\Phi} \mathcal{L}_n(\phi,\bfvtheta) +
\gamma J(\phi),
\end{equation}
where $\mathcal{L}_n$ is the loss function for observed data and
$J(\phi)$ imposes some type of penalty on the complexity of $\phi$.
Our approach is to replace $\mathcal{L}_n$ with the \citet{Gehan65}
loss function [\citet{Jin03}] and model $\phi$ using penalized
regression splines; our focus is to build risk prediction scores. To
minimize the penalized loss function \eqref{eq:pen.loss}, the insight
into the optimization procedure is due, in part, to \citet{Koenker94},
who noted that the optimization problem in quantile smoothing splines
can be solved by $L_1$-type linear programming techniques and proposed
an interior point algorithm for the problem. \citet{Li07} built on
this idea to propose an entirely different path-finding algorithm for
more general nonparametric quantile regression models. Along similar
lines, when $J(\phi)$ is taken as a $L_1$ norm as in penalized
regression splines [\citet{Ruppert97}], the optimization problem of
\eqref{eq:pen.loss} is essentially an $L_1$ loss plus $L_1$ penalty
problem, and can also be solved by $L_1$-type linear programming
techniques, which will be exploited in our approach to the optimization
problem. Once the basic spline framework is adopted, we show that our
estimator can be generalized through additive models for $q>1$ and
variable selection in the linear component. The additive structure of
nonlinear components [\citet{Hastie90}] is adopted to further alleviate
the issue of curse of dimensionality. To the best of our knowledge,
there is no similar work in the partly linear or partly additive model
for censored or uncensored data using Cox or AFT models, and we are the
first to conduct systematic investigation on the impact of misspecified
nonlinear effects on prediction and feature selection using AFT models
for high-dimensional data.

More recently, \citet{Chen05} proposed stratified rank estimation for
Model \eqref{eq:plm} and \citet{Johnson09} proposed a regularized
extension. However, their stratified methods are fundamentally
different from ours in several aspects. First and foremost, the
stratified estimators do not provide an estimate of the nonlinear
effect of the stratifying variable, namely,~$\hat\phi(X)$, and,
hence, the lasso extension proposed by \citet{Johnson09} focused on
variable selection only. It is evident that $\hat\phi(X)$ plays an
important role in prediction; since the stratified estimators in \citet
{Johnson09} can only use $\widehat{\bfvtheta}^{\mathrm{T}} \bfZ$ for
prediction, their performance suffers, which will be shown in our
numerical studies. By contrast, our approach provides an estimate of
$\phi(X)$, which in turn can be used to improve prediction
performance. Second, the numerical algorithm proposed in \citet
{Johnson09} can only handle the case of $d<n$ and their numerical
studies are limited to such cases, whereas we here investigate the
high-dimensional settings with $d>n$. Third, as will be shown in our
numerical results, our proposed method outperforms the stratified
estimators in feature selection as well.

The rest of the article is organized as follows. In Section \ref
{sec:methods} we present the details of the methodology. In Section
\ref{sec:sim} we investigate the operation characteristics of the
proposed approach through simulation studies. In Section~\ref
{sec:analysis} we analyze the prostate cancer study and provide answers
to the research questions of interest. We conclude this article with
some discussion remarks in Section \ref{sec:discussion}.

\section{Methodology}\label{sec:methods}

\subsection{Regression splines in partly linear AFT model}\label{sec:rs}
We first consider a~simplified case for the partly linear AFT model
\eqref{eq:plm}, where $\bfX_i$ is assumed to be univariate, that is,
$q=1$ and $\bfX_i\equiv X_i$, and then Model \eqref{eq:plm} reduces to
%
\begin{equation}\label{eq:plm.simp}
T_i= \phi(X_i) + \bfvtheta^{\mathrm{T}} \bfZ_i + \varepsilon_i.
\end{equation}
Let $\mathbb{B}(x)=\{B_1(x), \ldots, B_M(x)\}^{\mathrm{T}}$ ($M\le n$) be
a set of basis functions. We use a regression spline model for $\phi
(\cdot)$, which asserts that $\phi(x)=\mathbb{B}(x)^{\mathrm{T}}\bfbeta
$, for some $\bfbeta\in\Re^M$. Popular bases include $B$-splines,
natural splines and truncated power series basis [\citet{Ruppert03}].
As explained in Section \ref{sec:prs}, we will use the truncated power
series basis of degree $p$ without the intercept term, that is,
$\mathbb{B}(x)= \{x,\ldots,x^p,(x-\kappa_1)^p_+,\ldots, (x-\kappa
_r)^p_+ \}^{\mathrm{T}}$, where $(\kappa_1,\ldots,\kappa_r)$ denotes a
set of $r$ knots, and $(u)_+=uI(u\geq0)$. Hence, $M=p+r$. Throughout,
we use equally spaced percentiles as knots and set $p=3$, that is, the
cubic splines, unless otherwise noted. Let $\bftheta\equiv(\bfbeta
,\bfvtheta)$ denote the parameters of interest. Then, define $\widehat
\bftheta_{RS}\equiv(\widehat\bfbeta,\widehat\bfvtheta) =
\operatorname{argmin}_{\bfbeta,\bfvtheta} \mathcal{L}_n(\bfbeta,\bfvtheta)$,
where
%
\begin{equation}\label{eq:Gehan}
\mathcal{L}_n(\bfbeta,\bfvtheta) = n^{-2}\sum_{i=1}^n \sum_{j=1}^n
\delta_i (e_i - e_j)_{-}
\end{equation}
with $e_i = \Ttil_i - \bfbeta^{\mathrm{T}}\mathbb{B}(X_i)- \bfvtheta
^{\mathrm{T}}\bfZ_i$ and $c_{-}=\max(0,-c)$. Because Model \eqref
{eq:plm.simp} has been ``linearized,'' we can apply existing rank-based
estimation techniques for the usual linear AFT models. In particular,
\citet{Jin03} noted that the minimizer of $\mathcal{L}_n(\bfbeta
,\bfvtheta)$ is also the minimizer of
\[
\sum_{i=1}^n \sum_{j=1}^n \delta_i |e_i - e_j| + \Biggl| \zeta- (\bfbeta
^{\mathrm{T}},\bfvtheta^{\mathrm{T}})\sum_{k=1}^{n}\sum_{l=1}^n \delta_k
D_{lk} \Biggr|
\]
for a large constant $\zeta$, where $D_{lk}= \{\mathbb{B}(X_l)^{\mathrm{T}},\bfZ_l^{\mathrm{T}} \}^{\mathrm{T}} - \{\mathbb{B}(X_k)^{\mathrm{T}},\bfZ_k^{\mathrm{T}} \}^{\mathrm{T}}$. Evidently, the minimizer of this new loss function may
be viewed as the solution to a $L_1$ regression of a pseudo response
vector $\bfV=(V_1,\ldots,V_S)^{\mathrm{T}}$ ($S\times1$) on a pseudo
design matrix $\bfW=(\bfW_1,\ldots,\bfW_S)^{\mathrm{T}}$ ($S\times
(M+d))$. It can be readily shown that $\bfV$ is of the form $ \{\delta
_i(\Ttil_i-\Ttil_j),\ldots,\zeta\}^{\mathrm{T}}$ and $\bfW$ is of the
form $ (\delta_i D_{ij}, \ldots,\sum_{k=1}^{n}\sum_{l=1}^n\delta_k
D_{lk} )^{\mathrm{T}}$, where $\delta_i(\Ttil_i-\Ttil_j)$ and $\delta_i
D_{ij}^{\mathrm{T}}$ go through all $i$ and $j$ with $\delta_i=1$, and,
hence, $S$~denotes the number of pseudo observations in $\bfV$.
Consequently, we have
%
\begin{equation}\label{eq:regression.spline}
\widehat\bftheta_{\mathrm{RS}} = \operatorname{argmin}\limits_{\bfbeta,\bfvtheta} \sum
_{s=1}^S | V_s - \bftheta^{\mathrm{T}}\bfW_s |.
\end{equation}
The fact that $\widehat\bftheta_{\mathrm{RS}}$ can be written as the $L_1$
regression estimate facilitates the numerical techniques, which will be
used for our subsequent estimators.

\subsection{Penalized regression splines in partly linear AFT models}
\label{sec:prs} When regression splines are used to model nonlinear
covariates effects, it is crucial to choose the optimal number and
location of knots $(\kappa_1,\ldots,\kappa_r)$. It is well known that
too many knots may lead to overfitting, whereas too few may not be
sufficient to capture nonlinear effects [\citet{Ruppert03}]. The
penalized regression spline regression approach [\citet{Eilers96};
\citet{Ruppert97}; \citet{LiY08}; \citet{Claeskens09}]
handles this problem by starting with a~very large number of knots and
applying regularization to avoid overfitting. In addition, a penalized
regression spline with $L_1$ penalty corresponds to a~Bayesian model
with double exponential or Laplace priors and is known to be able to
accommodate large jumps when using the truncated polynomial basis
functions [\citet{Ruppert97}]. While the truncated power series
basis is often used for penalized regression spline [\citet{Ruppert97}], one can use other bases such as B-splines basis in
penalized regression spline models and the results should not differ as
long as two sets of bases span the same space of functions [\citet{LiY08}]. We adopt the $L_1$ penalty and consider the penalized
regression spline estimator
%
\begin{eqnarray}\label{eq:penalized.spline}
\widehat\bftheta_{\mathrm{PRS}}(\gamma) = \operatorname{argmin}\limits_{\bfbeta
,\bfvtheta} \Biggl\{
\mathcal{L}_n(\bfbeta, \bfvtheta) + \gamma\sum_{m=p+1}^M |\beta
_m| \Biggr\},
\end{eqnarray}
referred to as the partly linear AFT estimator, where $\gamma$ is a
regularization parameter and is used to achieve the goal of knot
selection. Using the~$L_1$ loss function in \eqref
{eq:regression.spline} and a data augmentation technique for
regularized~$L_1$ regression, $\bftheta_{\mathrm{PRS}}(\gamma)$ may be
found easily for a given $\gamma$. Namely, define $\bfV^{\ast}=(\bfV
^{\mathrm{T}},{\mathbf{0}}_r^{\mathrm{T}})^{\mathrm{T}}$, $\bfW^{\ast}=[\bfW^{\mathrm{T}},({\mathbf{0}}_{r\times p},\bfD_r, {\mathbf{0}}_{r\times d})^{\mathrm{T}}]^{\mathrm{T}}$, and $\bfD_r=\gamma I_r$, where ${\mathbf{0}}_r$ is a $r$-vector of
zeros, ${\mathbf{0}}_{r\times p}$ (${\mathbf{0}}_{r\times d}$) is a $r\times p$
($r\times d$) matrix of zeros and~$I_r$ an $r$-dimensional identity
matrix. Then, $\widehat\bftheta_{\mathrm{PRS}}(\gamma)$ is found through
the~$L_1$ regression of $\bfV^{\ast}$ on $\bfW^{\ast}$. $\gamma$
can be selected through cross-validation or generalized
cross-validation [\citet{Ruppert03}].

\subsection{Variable selection and prediction in partly linear AFT models}\label{sec:selection}
Finally, we consider variable selection for the high-dimensional
features ($\bfZ$) in the partly linear AFT model \eqref{eq:plm.simp}
by extending the penalized regression spline estimator $\widehat
\bftheta_{\mathrm{PRS}}(\gamma)$. Let $\lambda$ be another regularization
parameter and consider the minimizer to the $L_1$ regularized loss function
%
\begin{equation}\label{eq:lasso}
\widehat\bftheta_{\mathrm{PRS}(1)}(\gamma,\lambda) = \operatorname{argmin}\limits_{\bfbeta,\bfvtheta} \Biggl\{ \mathcal{L}_n(\bfbeta, \bfvtheta)
+ \gamma\sum_{m=p+1}^M |\beta_m| + \lambda\sum_{j=1}^d |\vartheta
_j| \Biggr\},
\end{equation}
which is also referred to as the lasso partly linear AFT model
estimator. The data augmentation scheme used in Section \ref{sec:prs}
applies to the regularized estimator in \eqref{eq:lasso} as well.
Define the pseudo response vector $\bfV^{\dagger}=(\bfV^{\mathrm{T}},
{\mathbf{0}}_{r+d}^{\mathrm{T}})^{\mathrm{T}}$ and the pseudo design matrix
\[
\bfW^{\dagger}= \left[ \bfW^{\mathrm{T}}, \pmatrix{
{\mathbf{0}}_{r\times p}&\gamma I_r & {\mathbf{0}}_{r\times d}
\cr
{\mathbf{0}}_{d\times p} & {\mathbf{0}}_{d\times r} &\operatorname{diag}(\lambda
_1,\ldots,\lambda_d)
}^{\mathrm{T}}
\right]^{\mathrm{T}}.
\]
For fixed $\gamma$ and $\lambda$, the estimate is computed as the
$L_1$ regression estimate of $\bfV^{\dagger}$ on $\bfW^{\dagger}$.
To select $\gamma$ and $\lambda$, we can use two approaches, namely,
the cross-validation (CV) and the generalized cross-validation (GCV)
[\citet{Tibshirani97}; \citet{Cai09}]. The $K$-fold CV approach chooses the
values of $\gamma$ and $\lambda$ that maximize the Gehan loss
function \eqref{eq:Gehan}. The GCV approach chooses the values of
$\gamma$ and $\lambda$ that maximize the criteria, $\mathcal
{L}_n(\bfbeta,\bfvtheta)/(1-d_{\gamma,\lambda}/n)^2$, where $n$ is
the number of observations and $d_{\gamma,\lambda}$ is the number of
nonzero estimated coefficients for the basis functions ($\mathbb
{B}(X)$) and linear predictors ($Z$), that is, the number of nonzero
estimates in $(\widehat\bfbeta,\widehat\bfvtheta)$. Note that
$d_{\gamma,\lambda}$ depends on $\gamma$ and $\lambda$. Once
$\bftheta_{\mathrm{PRS}(1)}$ is obtained, one can build prediction scores
as $\widehat\phi(X) + \widehat\bfvtheta^{\mathrm{T}} \bfZ$.

\subsection{Extension to additive partly linear AFT models} \label{sec:additive}
When $\bfX_i$ is of $q$-dimension ($q>1$) in the partly linear model
(\ref{eq:plm}), estimation is more difficult due to the issue of curse
of dimensionality, even when $q$ is moderately large and in the absence
of censoring. For our partly linear AFT model, we propose to use an
additive structure for $\phi$ to further alleviate the problem,
namely, an additive partly linear AFT model,
%
\begin{equation}\label{eq:pam}
T_i= \sum_{j=1}^q \phi_j\bigl(X_i^{(j)}\bigr) + \bfvtheta^{\mathrm{T}} \bfZ_i +
\varepsilon_i,
\end{equation}
where $\phi_j$'s $(j=1,\ldots,q)$ are unknown functions. Similar to
what is discussed in Section \ref{sec:prs}, penalized regression
splines can be used for the additive partly linear model to conduct
knot selection for each nonlinear effect, $\phi_j(X_i^{(j)})$
($j=1,\ldots, q$). The variable selection for $\bfZ$ as discussed in
Section~\ref{sec:selection} can also be extended to this additive
partly linear AFT model. When $q$ is large and it is also of interest
to conduct feature selection among $q$ additive nonlinear effects, one
can modify the regularization term for $\bfbeta$ in the loss
functions~\eqref{eq:penalized.spline} and \eqref{eq:lasso}; specifically, one
can regularize all $\bfbeta$, that is, $\gamma\sum_{m=1}^M |\beta
_m|$, as opposed to regularizing only the terms that correspond to the
set of jumps in the $p$th derivative, that is, $\gamma\sum_{m=p+1}^M
|\beta_m|$. Similarly, we can modify the data augmentation scheme to
obtain the parameter estimates for these models.

\subsection{Numerical implementation for high-dimensional data}\label{sec:implementation}
In Sec-\break{}tions~\mbox{\ref{sec:rs}--\ref{sec:additive}}~the parameters are
estimated using $L_1$ regression models through a data augmentation
scheme such as \eqref{eq:regression.spline}, which can be readily
implemented using the \texttt{quantreg} package in \texttt{R}. While this
algorithm works well when the total number of parameters is small
relative to the sample size, it becomes very slow and starts to fail as
the number of parameters gets close to or greater than the effective
sample size after accounting for censoring. As an alternative, we
extended a numerical algorithm developed for efficient computation of
rank estimates for AFT models [\citet{Conrad10}] to compute the
proposed estimators, in particular, the estimator in \eqref{eq:lasso}.
In essence, this method approximates a $L_1$ regularized loss function
with a smooth function and subsequently optimizes the smoothed
objective function using a Limited-Broyden--Fletcher--Goldfarb--Shanno
(L-BFGS) algorithm [\citet{Nocedal06}], which is implemented in \texttt{Matlab}. This method speeds up the computation substantially and can
handle the case of high-dimensional data. We have compared these two
algorithms and they give very similar results when both are applicable,
that is, $\bfZ$ is of low dimension.

\section{Simulation studies}\label{sec:sim}
We conducted extensive simulation studies to evaluate the operating
characteristics of the proposed models including estimation, feature
selection and, most importantly, prediction, in comparison with several
existing models.

\subsection{Estimation}\label{sec:sim-est}
We considered a case of single $Z_i$ and single $X_i$, that is, Model
\eqref{eq:plm.simp}, and focused on the estimation of the regression
coefficient $\vartheta$ and its sampling variance. In this setup, no
feature selection is involved. To facilitate comparisons, our
simulation study details were adapted from those given by \citet
{Chen05} and \citet{Johnson09}. The random variable $Z_i$ was generated
from a standard normal distribution, and $X_i$ was generated through
$X_i=0.25Z_{i}+U_i$, where $U_i$ follows a uniform distribution $\operatorname{Un}(-5,5)$ and completely independent of all other random variables.
In Model \eqref{eq:plm.simp} we let $\vartheta=1$ and $\varepsilon
_i\sim N(0,1)$ and mutually independent of $(X_i,Z_i)$. We considered
linear and quadratic effects, that is, $\phi(X_i)=2X_i$ and $\phi
(X_i)=X_i^2$, respectively. Finally, censoring random variables were
simulated through $C_i=\phi(X_i) + Z_i\vartheta+ U^{\ast}_i,$ where
$U_i^{\ast}$ follows $\operatorname{Un}(0,1)$. As a~result, the proportion of
censored outcomes ranges from 20\% to 30\%. We compared several
estimators, the partly linear AFT model (PL-AFT) with~$r$ knots ($r=2$
and 4), which was fit using the loss function \eqref
{eq:penalized.spline}, the stratified estimator in \citet{Chen05}
(S$_K$-AFT) where $K$ denotes the number of strata, the standard linear
AFT model with both $X_i$ and $Z_i$ modeled linearly (AFT), and an AFT
model with true $\phi$ plugged in (AFT-$\phi$). Two sample sizes were
used, $n=50$ and $n=100$.

\begin{table}
\caption{Simulation results for parameter estimation ($\hat\vartheta
$) based on 200 Monte Carlo data sets, where $\vartheta=1$}\label{tab:simu1}
\begin{tabular*}{\tablewidth}{@{\extracolsep{\fill}}ld{2.0}ccd{2.0}d{4.0}d{3.0}@{}}
\hline
&\multicolumn{3}{c}{$\bolds{\phi(X)=2X}$} & \multicolumn{3}{c@{}}{$\bolds{\phi(X)=2X^2}$}\\[-5pt]
&\multicolumn{3}{c}{\hrulefill}&\multicolumn{3}{c@{}}{\hrulefill}\\
& \multicolumn{1}{c}{\textbf{Bias}} & \textbf{SD} & \textbf{MSE} &\multicolumn{1}{c}{\textbf{Bias}} & \multicolumn{1}{c}{\textbf{SD}} & \multicolumn{1}{c@{}}{\textbf{MSE}}\\
\hline
& \multicolumn{6}{c@{}}{$n=50$}\\
PL-AFT ($r=2$) &-12 & 159 & 25 & -2 & 166 & 28 \\
PL-AFT ($r=4$) &-10 & 159 & 25 & -1 & 168 & 28 \\
S$_5$-AFT & 95 & 288 & 92 &-65 & 436 & 195\\
S$_{10}$-AFT & 28 & 223 & 50 &-43 & 299 & 91 \\
S$_{25}$-AFT & 31 & 303 & 93 &-38 & 381 & 146\\
AFT & -4 & 153 & 23 & 21 &1\mbox{,}214 &1\mbox{,}475\\
AFT-$\phi$ & -7 & 154 & 24 & -5 & 158 & 25 \\[3pt]
&\multicolumn{6}{c@{}}{$n=100$} \\
PL-AFT ($r=2$) & -9 & 113 & 13 & -2 & 115 & 13 \\
PL-AFT ($r=4$) & -9 & 113 & 13 & -1 & 115 & 13 \\
S$_{10}$-AFT & 44 & 163 & 29 &-23 & 210 & 45 \\
S$_{25}$-AFT & 1 & 157 & 25 & -9 & 185 & 34 \\
S$_{50}$-AFT & -7 & 193 & 37 & 8 & 209 & 44 \\
AFT & -8 & 113 & 13 & 71 & 755 &575 \\
AFT-$\phi$ & -9 & 113 & 13 & -2 & 111 & 12 \\[5pt]
Range of SEs &\multicolumn{1}{c}{8--21}& NA &1--12&\multicolumn{1}{c}{8--86}&\multicolumn{1}{c}{NA} &\multicolumn{1}{c@{}}{1--209}\\
\hline
\end{tabular*}
\legend{PL-AFT,
partly linear AFT model with $r$ knots; S$_K$-AFT, stratified AFT
estimator with $K$ strata; AFT, standard linear AFT model with both
$X_i$ and $Z_i$ modeled linearly; and AFT-$\phi$, AFT model with true
$\phi$ plugged in. Range of SEs, the range of SEs for the
corresponding performance measure in each column. NA, SE of a
performance measure cannot be computed for SD. All numbers are
multiplied by 1,000.}
\end{table}

Our simulation results show that the CV and GCV methods give similar
results, so we report only the results using GCV. Table \ref
{tab:simu1} summarizes the mean bias, standard deviation (SD) and mean
squared error (MSE) of $\hat\vartheta$ over 200 Monte Carlo data
sets, and it also provides the range of standard errors for the
performance measure in each column, where all numbers are multiplied by
1,000. In all cases, the proposed partly linear AFT estimator
outperforms the stratified estimators as well as the standard AFT
estimator in terms of MSE, and its performance is comparable to that of
the estimator using the true $\phi$. The number of knots has little
impact on the performance of our proposed estimator. The standard
linear AFT estimator exhibits the largest bias and MSE when $\phi$ is
not linear, indicating that it is important to adjust for the nonlinear
effect of $X$ even when one is only interested in the effect of $Z$.
While the stratification step in the S$_K$-AFT method results in
reduced bias when the number of strata is large, it has larger SD and
MSE compared to PL-AFT. Furthermore, in the settings of our interest,
no method has been proposed for choosing $K$ in the S$_K$-AFT method,
which is not obvious either, leading to a further shortcoming of this
method over the others.

\subsection{Feature selection}\label{sec:sim-pred}
In our second set of simulation studies, we focused on simultaneous
estimation and feature selection for $\bfZ_i$ as well as prediction.
The regression function still consisted of a nonlinear effect of a
single covariate $X_i$, but we increased the dimension of the linear
predictors ($\bfZ_i$) to $d=8$. $\bfZ_i$~were generated from a
multivariate normal with a mean equal to $0_d$ and $(j,k)$th element of
the covariance matrix equal to $\rho^{|j-k|}$ ($\rho=0,0.5,0.9$). The
covariate $X_i$ was generated through
$X_i=0.5Z_{1i}+0.5Z_{2i}+0.5Z_{3i}+U_i$, where $U_i$ is $\operatorname{Un}(-1,1)$ and independent of all other random variables. This
corresponds to a case where $Z_{1}$ and $Z_{2}$ have both direct and
indirect effects through $X$ on the outcome, whereas~$Z_3$ has only an
indirect effect on the outcome. The true regression coefficients for
$\bfZ$ are set to $\bfvtheta=(\Delta, \Delta, 0, 0, 0, \Delta, 0,
0)'$, where $\Delta=1$ and 0.5 represent a~strong signal (effect size)
and a weak signal (effect size), respectively. In this case, the three
important covariates (namely, $Z_1$, $Z_2$ and $Z_6$) can potentially
be highly correlated. The effect of $X_i$ was generated from $\phi
(X_i)=(0.2*X_i+0.5*X_i^2+0.15*X_i^3)I(X_i\geq0)+(0.05*X_i)I(X_i<0)$,
where $I(\cdot)$ is the indicator function. This setup mimics a
practical setting where the effect of the clinical variable ($X$) on
the outcome is ignorable when $X$ is less than a threshold level
($X=0$); but as $X$ increases past the threshold level, its effect
becomes appreciable. The log survival time $T_i$ was then generated
using equation \eqref{eq:plm.simp}, where $\varepsilon_i$ follows
$N(0,1)$ and is mutually independent of $(X_i,\bfZ_i)$. The censoring
random variable was simulated according to the rule, $C_i=\phi(X_i) +
\bfvtheta^T\bfZ_i + U^{\ast}_i,$ where $U_i^{\ast}$ follows the
uniform distribution $\operatorname{Un}(0,6)$. The resulting proportion of
censoring ranges from 20\% to 30\%.

We compared six models: (1) the lasso partly linear AFT model
(Lasso-PL) with $r=6$ which was fit using the loss function \eqref
{eq:lasso}; (2) the lasso stratified model (Lasso-S$_K$) [\citet{Johnson09}] where $K$ denotes the number of strata; (3) the lasso
linear AFT model assuming a linear effect for both~$X_i$ and $\bfZ_i$
(Lasso-L); (4) the standard linear AFT model (AFT); (5) the lasso
linear Cox PH model assuming a linear effect for both $X_i$ and $\bfZ
_i$ (Lasso--Cox) [\citet{Tibshirani97}; \citet{Goeman10}]; and (6) the so-called
oracle partly linear model (Oracle) with $\vartheta_3$, $\vartheta
_4$, $\vartheta_5$, $\vartheta_7$ and $\vartheta_8$ fixed at 0 and
$r=6$ for the penalized splines. We are not aware of any existing Cox
PH model that can handle both nonlinear covariate effects and feature
selection in high-dimensional data. Since the data were generated under
a true AFT model and the PH assumption underlying the Cox model is
violated, we are primarily interested in feature selection when
comparing the Lasso--Cox model. The oracle model, while unavailable in
practice, may serve as an optimal benchmark for the purpose of
comparisons. In each instance of regularized methods, GCV was used to
tune the regularization parameters, $\lambda$ and/or $\gamma$.

In each simulation run, a training sample of size $n=125$ and a testing
sample of size $10n$ were generated. To evaluate parameter estimation,
we monitored the sum of squared errors (SSE) for $\widehat\bfvtheta$
defined as $(\widehat\bfvtheta-\bfvtheta)^{\mathrm{T}}(\widehat\bfvtheta
-\bfvtheta)$. To evaluate feature selection, we monitor the proportion
of zero coefficients being set to zero ($P_C\equiv\sum
_{i=1}^dI(\widehat\vartheta_i=0)I(\vartheta_i=0)/\sum
_{i=1}^dI(\vartheta_i=0)$), for which~1 is the optimal value, and the
proportion of nonzero coefficients being set to zero ($P_I\equiv\sum
_{i=1}^dI(\widehat\vartheta_i=0)I(\vartheta_i\neq0)/\sum
_{i=1}^dI(\vartheta_i\neq0)$), for which~0 is the optimal value. To
assess the prediction performance, we considered two mean squared
prediction errors, $\mathrm{MSPE}_1\equiv(10n)^{-1} \sum_{j=1}^{10n}
[\hat\phi(X_j)-\phi(X_j)+(\widehat\bfvtheta- \bfvtheta)^{\mathrm{T}}\bfZ_j ]^2$, and $\mathrm{MSPE}_2\equiv(10n)^{-1}\sum_{j=1}^{10n}
[(\widehat\bfvtheta- \bfvtheta)^{\mathrm{T}}\bfZ_j ]^2$, where $j$ goes
through the observations in the testing sample. $\mathrm{MSPE}_1$ is the
squared prediction error using both nonlinear and linear components in
Model (\ref{eq:plm.simp}), and $\mathrm{MSPE}_2$ is the squared
prediction error using only linear components in Model (\ref
{eq:plm.simp}). For AFT models, $\mathrm{MSPE}_1$ and $\mathrm{MSPE}_2$
can be considered as metrics of prediction performance on the
log-transformed scale. Note that the stratified Lasso model does not
provide an estimate of $\phi(X)$, so $\mathrm{MSPE}_1$ is not
applicable for Lasso-S$_K$. For each simulation setting, the
performance measures were averaged over 400 Monte Carlo data sets. For
the performance measure in each column, the range of standard errors
was computed.

Our simulation results are summarized in Table \ref{tab:low-dim}.
First, the performance of the standard linear AFT model (AFT) is not
satisfactory in terms of both prediction and feature selections. We now
restrict the discussion to the regularized estimators. In all cases,
our Lasso-PL estimator exhibits lowest SSE, MSPE$_1$ and MSPE$_2$ among
regularized estimators; in particular, its MSPE$_1$ and MSPE$_2$ are
comparable to that of the Oracle estimator and are substantially lower
than other regularized estimators. In terms of feature selection,
Lasso-PL, Lasso-L and Lasso--Cox correctly identify the majority of the
regression coefficients that are zero ($P_C$); Lasso-PL has
higher~$P_C$ than \mbox{Lasso-L} when $\rho=0$ or 0.5 and their~$P_C$'s are
comparable in the presence of high correlation ($\rho=0.9$); and
Lasso-L has considerably higher~$P_C$ than Lasso--Cox in all cases. By
comparison, the lasso stratified models \mbox{(Lasso-S$_K$)} only identify
less than 30\% of true zeros in some cases and roughly half of the true
zeros in the rest of the cases. When there is no correlation and the
signal is strong, all Lasso estimators successfully avoid setting
nonzero coefficients to zero, that is, $P_I$ equal to or close to 0.
However, as the correlation gets stronger, $P_I$ increases for all
estimators to various degrees. When $\rho=0.9$, $P_I$ becomes
appreciable for Lasso-L, whereas it remains moderate for Lasso-PL.

\begin{table}
\tabcolsep=0pt
\caption{Simulation results for evaluating feature selection and
prediction performance based on 400 Monte Carlo data sets, where
$n=125$ and $d=8$}\label{tab:low-dim}
\begin{tabular*}{\tablewidth}{@{\extracolsep{\fill}}lcccccccccc@{}}
\hline
& \multicolumn{5}{c}{$\bolds{\Delta=1}$} & \multicolumn{5}{c@{}}{$\bolds{\Delta=0.5}$}\\[-5pt]
&\multicolumn{5}{c}{\hrulefill}&\multicolumn{5}{c@{}}{\hrulefill}\\
&\multicolumn{1}{c}{\textbf{SSE}} & \multicolumn{1}{c}{$\bolds{P_C}$} & \multicolumn{1}{c}{$\bolds{P_I}$}&
\multicolumn{1}{c}{\textbf{MSPE}$_{\mathbf{1}}$} &\multicolumn{1}{c}{\textbf{MSPE}$_{\mathbf{2}}$} & \multicolumn{1}{c}{\textbf{SSE}} &
\multicolumn{1}{c}{$\bolds{P_C}$} & \multicolumn{1}{c}{$\bolds{P_I}$} & \multicolumn{1}{c}{\textbf{MSPE}$_{\mathbf{1}}$} &
\multicolumn{1}{c@{}}{\textbf{MSPE}$_{\mathbf{2}}$}\\
\hline
&\multicolumn{10}{c@{}}{$\rho=0$}\\
Lasso-PL & \hphantom{00}8 &\hphantom{0,}734 &\hphantom{0}0 &\hphantom{0,}244 & \hphantom{0}67 & \hphantom{00}8 &\hphantom{0,}724 &\hphantom{00}0 &\hphantom{0,}237 & \hphantom{0}67\\
Lasso-S$_2$ & \hphantom{0}23 &\hphantom{0,}482 &\hphantom{0}0 &NA &186 & \hphantom{0}23 &\hphantom{0,}453 & \hphantom{00}1 &NA &185\\
Lasso-S$_4$ & \hphantom{0}16 &\hphantom{0,}582 &\hphantom{0}0 &NA &127 & \hphantom{0}15 &\hphantom{0,}565 & \hphantom{00}2 &NA &122\\
Lasso-S$_8$ & \hphantom{0}20 &\hphantom{0,}424 &\hphantom{0}0 &NA &161 & \hphantom{0}20 &\hphantom{0,}438 & \hphantom{00}8 &NA &159\\
Lasso-L & \hphantom{0}12 &\hphantom{0,}639 &\hphantom{0}0 &\hphantom{0,}997 &100 & \hphantom{0}12 &\hphantom{0,}611 &\hphantom{00}0 &\hphantom{0,}990 & \hphantom{0}99\\
Lasso--Cox & NA &\hphantom{0,}488 &\hphantom{0}0 &NA &NA & NA &\hphantom{0,}543
&\hphantom{0}17 &NA &NA \\
AFT & \hphantom{0}18 &\hphantom{0,00}0 &\hphantom{0}0 &\hphantom{0,}982 &142 & \hphantom{0}18 &\hphantom{000,}0 &\hphantom{00}0 &\hphantom{0,}982 &143\\
Oracle & \hphantom{00}4 &1,000&\hphantom{0}0 &\hphantom{0,}153 & \hphantom{0}29 & \hphantom{00}4 &1,000&\hphantom{00}0 &\hphantom{0,}207 & \hphantom{0}30\\[3pt]
&\multicolumn{10}{c@{}}{$\rho=0.5$}\\
Lasso-PL & \hphantom{0}11 &\hphantom{0,}767 &\hphantom{0}0 &\hphantom{0,}225 & \hphantom{0}74 & \hphantom{0}11 &\hphantom{0,}777 & \hphantom{00}2 &\hphantom{0,}296 & \hphantom{0}75\\
Lasso-S$_2$ & \hphantom{0}38 &\hphantom{0,}403 &\hphantom{0}0 &NA &341 & \hphantom{0}40 &\hphantom{0,}412 & \hphantom{00}8 &NA &353\\
Lasso-S$_4$ & \hphantom{0}21 &\hphantom{0,}569 &\hphantom{0}0 &NA &171 & \hphantom{0}20 &\hphantom{0,}599 & \hphantom{00}5 &NA &146\\
Lasso-S$_8$ & \hphantom{0}26 &\hphantom{0,}540 &\hphantom{0}0 &NA &218 & \hphantom{0}26 &\hphantom{0,}594 &\hphantom{0}15 &NA &204\\
Lasso-L & \hphantom{0}19 &\hphantom{0,}720 &\hphantom{0}0 &2,894&126 & \hphantom{0}19 &\hphantom{0,}748 &\hphantom{0}16 &2,943&121\\
Lasso--Cox & NA &\hphantom{0,}562 &\hphantom{0}0 &NA &NA & NA &\hphantom{0,}612 &\hphantom{0}14
&NA &NA \\
AFT & \hphantom{0}33 &\hphantom{0,00}0 &\hphantom{0}0 &2,839&212 & \hphantom{0}32 &\hphantom{0,00}0 &\hphantom{00}0 &2,878&202\\
Oracle & \hphantom{00}5 &1,000&\hphantom{0}0 &\hphantom{0,}175 & \hphantom{0}31 & \hphantom{00}5 &1,000&\hphantom{00}0 &\hphantom{0,}248 & \hphantom{0}32\\[3pt]
&\multicolumn{10}{c@{}}{$\rho=0.9$}\\
Lasso-PL & \hphantom{0}45 &\hphantom{0,}739 & \hphantom{0}2 &\hphantom{0,}373 &118 & \hphantom{0}39 &\hphantom{0,}758 &113 &\hphantom{0,}337 &130\\
Lasso-S$_2$ &126 &\hphantom{0,}502 &16 &NA &592 &106 &\hphantom{0,}500 &152 &NA &595\\
Lasso-S$_4$ & \hphantom{0}77 &\hphantom{0,}582 & \hphantom{0}4 &NA &184 & \hphantom{0}60 &\hphantom{0,}596 &124 &NA &170\\
Lasso-S$_8$ &118 &\hphantom{0,}236 & \hphantom{0}6 &NA &338 & \hphantom{0}96 &\hphantom{0,}424 &135 &NA &390\\
Lasso-L & \hphantom{0}92 &\hphantom{0,}751 &31 &6,571&245 & \hphantom{0}65 &\hphantom{0,}778 &270 &6,738&262\\
Lasso--Cox & NA &\hphantom{0,}596 & \hphantom{0}8 &NA &NA & NA
&\hphantom{0,}651 &153 &NA &NA \\
AFT &224 &\hphantom{0,00}0 &\hphantom{0}0 &6,483&337 &226 &\hphantom{0,00}0 &\hphantom{00}0 &6,612&354\\
Oracle & \hphantom{0}17 &1,000&\hphantom{0}0 &\hphantom{0,}320 & \hphantom{0}55 & \hphantom{0}17 &1,000&\hphantom{00}0 &\hphantom{0,}288 & \hphantom{0}54\\[5pt]
Range of SEs&0.1--8&0--24&0--5&8--76& 1--23&0.2--8&0--26&0--13&10--81&1--25\\
\hline
\end{tabular*}
\legend{Lasso-PL, Lasso partly linear AFT model;
Lasso-S$_K$, Lasso stratified model with $K$ strata; Lasso-L, Lasso
linear AFT model assuming a linear effect for both $X_i$ and $\bfZ_i$;
Lasso--Cox, Lasso linear Cox model assuming a linear effect for both
$X_i$ and $\bfZ_i$; AFT, standard AFT model assuming linear effects
for both $X_i$ and $\bfZ_i$ without regularization; and Oracle, oracle
partly linear model with zero coefficients being set to 0. $\Delta$,
effect size; SSE, sum of squared errors for $\widehat\bfvtheta$;
$P_C$, proportion of zero coefficients being set to zero; $P_I$,
proportion of nonzero coefficients being set to zero; $\mathrm{MSPE}_1$,
squared prediction error using both nonlinear and linear components;
and $\mathrm{MSPE}_2$, squared prediction error using only linear
components. Range of SEs, range of SEs for the corresponding
performance measure in each column. NA, a performance measure is not
applicable for an estimator. All numbers are multiplied by 1,000.}
\end{table}

\subsection{Prediction in the presence of high-dimensional data}\label{sec:sim-pred-high}

We conducted a third set of simulations to explore the impact of noise
levels on the prediction performance in the presence of
high-dimensional data (i.e., $d\geq n$), and compared four models,
namely, Lasso-PL, Lasso-S$_K$, Lasso-L and Lasso--Cox. We note that the
standard AFT model is not applicable for high-dimensional data. The
simulation setup paralleled that in Section \ref{sec:sim-pred}. The
differences are noted as follows. The sample size was fixed to $n=100$
and the number of linear predictors was $d\geq100$, and let $\vartheta
_1=\vartheta_{26}=\vartheta_{51}=\vartheta_{76}=1$ and all other
$\vartheta$'s be 0. Let $X=0.5Z_{10}+0.5Z_{35}+0.5Z_{60}+U_i$, where
$U_i$ follows $\operatorname{Un}(-1,1)$. Through these changes, we investigated
a case where the significant linear predictors ($Z$) are not highly
correlated. The censoring random variable was generated similar to that
in Section \ref{sec:sim-pred} with a different uniform distribution
such that the censoring probability is approximately $40\% $. Since
$\mathrm{MSPE}_1$ and $\mathrm{MSPE}_2$ are not applicable in the presence
of censoring in practice, we computed another metric of prediction
performance using the testing sample, namely, the $c$ statistic for
censored data, which measures the proportion of concordance pairs based
on observed and predicted outcomes and ranges between 0 and 1 with 1
indicating perfect prediction [\citet{Kattan03};
\citet{Kattan03a}; \citet{Steyerberg10}]. In particular, the
comparison with Lasso--Cox is focused on $c$ statistics. Again, for
Lasso-S$_K$, $\mathrm{MSPE}_1$ was not applicable and
$\widehat\bfvtheta^{\mathrm{T}}\bfZ_j$ was used to compute the $c$
statistic; for the performance measure in each column, the range of
standard errors was computed.

%
\begin{table}
\tabcolsep=0pt
\caption{Simulation results for evaluating prediction performance in
the presence of high-dimensional data based on 400 Monte Carlo data
sets, where $n=100$}\label{tab:high-dim}
\begin{tabular*}{\tablewidth}{@{\extracolsep{\fill}}lccccccccc@{}}
\hline
& \multicolumn{3}{c}{$\bolds{d=100}$} & \multicolumn{3}{c}{$\bolds{d=500}$}&
\multicolumn{3}{c@{}}{$\bolds{d=1\mbox{\textbf{,}}500}$} \\[-5pt]
&\multicolumn{3}{c}{\hrulefill}&\multicolumn{3}{c}{\hrulefill}&\multicolumn{3}{c@{}}{\hrulefill}\\
& \textbf{MSPE}$_{\mathbf{1}}$ &\textbf{MSPE}$_{\bolds{2}}$ & $\bolds{c}$ &\textbf{MSPE}$_{\bolds{1}}$ &\textbf{MSPE}$_{\mathbf{2}}$& $\bolds{c}$ &
\textbf{MSPE}$_{\mathbf{1}}$&\textbf{MSPE}$_{\mathbf{2}}$& $\bolds{c}$ \\
\hline
&\multicolumn{9}{c@{}}{$\rho=0$}\\
Lasso-PL & \hphantom{0,}412 &349 &860 &\hphantom{0,}989 & \hphantom{0,}897 &840 &1,685&1,543&796 \\
Lasso-S$_2$ & NA &676 &811 &NA &1,589 &768 &NA &2,310&711 \\
Lasso-S$_4$ & NA &560 &812 &NA &1,428 &780 &NA &2,182&718 \\
Lasso-S$_8$ & NA &529 &811 &NA &1,454 &775 &NA &2,208&716 \\
Lasso-L & 1,441&568 &829 &2,752&1,666 &784 &3,719&2,496&697 \\
Lasso--Cox & NA &NA &798 &NA &NA &749 &NA &NA &684 \\[3pt]
&\multicolumn{9}{c@{}}{$\rho=0.5$}\\
Lasso-PL & \hphantom{0,}389 &330 &860 &1,034& \hphantom{0,}937 &839 &1,659&1,518&797 \\
Lasso-S$_2$ & NA &637 &810 &NA &1,653 &766 &NA &2,270&716 \\
Lasso-S$_4$ & NA &525 &812 &NA &1,472 &777 &NA &2,152&725 \\
Lasso-S$_8$ & NA &491 &811 &NA &1,512 &774 &NA &2,196&721 \\
Lasso-L & 1,418&550 &829 &2,803&1,720 &781 &3,703&2,513&701 \\
Lasso--Cox & NA &NA &799 &NA &NA &749 &NA &NA &690 \\[3pt]
&\multicolumn{9}{c@{}}{$\rho=0.9$}\\
Lasso-PL & \hphantom{0,}387 &328 &875 &1,084&1,124 &852 &1,795&1,909&811 \\
Lasso-S$_2$ & NA &529 &841 &NA &1,314 &815 &NA &2,059&769 \\
Lasso-S$_4$ & NA &474 &842 &NA &1,422 &812 &NA &2,253&759 \\
Lasso-S$_8$ & NA &455 &841 &NA &1,618 &805 &NA &2,473&744 \\
Lasso-L & 1,476&480 &852 &2,274&1,152 &836 &3,179&1,849&802 \\
Lasso--Cox & NA &NA &840 &NA &NA &825 &NA &NA &796 \\[5pt]
Range of SEs&9--20 &8--23&0.6--2&32--56&32--52 &1--4&47--61&47--57&2--5\\
\hline
\end{tabular*}
\legend{Lasso-PL, Lasso partly linear AFT model;
Lasso-S$_K$, Lasso stratified model with $K$ strata; \mbox{Lasso-L}, Lasso
linear AFT model assuming a linear effect for both $X_i$ and $\bfZ_i$;
and Lasso--Cox, Lasso linear Cox model assuming a linear effect for both
$X_i$ and $\bfZ_i$. $\mathrm{MSPE}_1$, the squared prediction error
using both nonlinear and linear components; $\mathrm{MSPE}_2$, the
squared prediction error using only linear components; and $c$, the
$c$-statistic for censored data. Range of SEs, range of SEs for the
corresponding performance measure in each column. NA, a performance
measure is not applicable for a estimator. All numbers are multiplied
by 1,000.}
\vspace*{6pt}
\end{table}

Table \ref{tab:high-dim} summarizes the prediction performance for
$d=100$, $d=500$ and $d=1\mbox{,}500$ over 400 Monte Carlo data sets. In the
presence of high-dimensional data, Table \ref{tab:high-dim} shows that
the proposed Lasso-PL always achieves the best prediction performance
in terms of the $c$ statistic as well as MSPE$_1$ and MSPE$_2$, and
Lasso--Cox always has lower $c$ than Lasso-PL and \mbox{Lasso-L}. By and large,
the prediction performance of Lasso-S$_K$ is comparable to that of
Lasso-L and is considerably worse than Lasso-PL in all cases, and, in
particular, the absence of the estimated nonlinear effect in $X$ leads
to substantial loss in the $c$ statistic. While Lasso-PL estimates the
nonlinear effect of $X$ well in all cases, the prediction error due to
the linear predictors (MSPE$_2$) starts to dominate as $d$ increases.
Since all significant predictors are in the first 100 predictors, the
cases of $d=1\mbox{,}500$ and $d=500$ simply add 1,100 and 400 noise predictors,
respectively, compared to the case of $d=100$. Our results indicate
that as the noise level increases the prediction performance
deteriorates for all models. For Lasso-L models, the prediction error
due to misspecified nonlinear effect of $X$ remains substantial in all
cases. In this setup, when correlation is weak or moderate ($\rho=0$
or $0.5$), the impact of correlation on prediction performance is
moderate, in particular, in terms of $c$; however, as correlation
becomes very strong ($\rho=0.9$), the prediction performance improves
considerably in terms of $c$ for all methods.

We performed additional simulations for a higher censoring rate, $60\%
$, and for different regression coefficient values, for example,
$\vartheta_1=\vartheta_{2}=\vartheta_{3}=\vartheta_{50}=1$ and all
other $\vartheta$'s set to 0, that is, the first three significant
predictors are highly correlated. Under all scenarios, the results on
comparisons between different models remain the same, but the
prediction performance worsens as the censoring rate increases.

In summary, the proposed lasso partly linear AFT model achieves best
performance in all three areas: estimation, feature selection and
prediction. While the lasso stratified estimator performs reasonably
well in estimation, its performance in feature selection and prediction
is not satisfactory. When a covariate effect is nonlinear, the
performance of Lasso-L worsens, and the deterioration can be
substantial in terms of prediction. When the PH assumption does not
hold, the performance of Lasso--Cox is considerably worse than Lasso-L.
Furthermore, if prediction is of primary interest, our results suggest
that it is advantageous to build prediction scores using data with less
noise variables.

\section{Data analysis: The prostate cancer study}\label{sec:analysis}
We analyzed the data from the prostate cancer study, which included 78
patients. The outcome of interest is time to prostate cancer
recurrence, which starts on the day of prostatectomy and is subject to
censoring; the observed survival time ranges from 2 months to 160
months and the censoring rate is 57.7\%. In the data analysis, the
log-transformed survival time was used to fit AFT models. Gene
expression data using 1,536 probes and two clinical variables (PSA and
gleason score) were measured from samples collected at the baseline
(i.e., right after the surgery) and were used in our analysis. Since
replicate RNA samples were collected and measured from some subjects,
we averaged the gene expression data over multiple RNA samples from a
same subject before subsequent analysis. The gleason score in this data
set ranges only between 5 and 9 and 91\% of patients had a score of
either 6 or 7; combining this with suggestions from the investigators,
the total gleason score was dichotomized as $\geq7$ or not.

Before the data analysis, all gene expression measurements were
preprocessed and standardized to have mean 0 and unit standard
deviation. Subsequently, Cox PH models were fit for each individual
probe and all probes were then ranked according to their score test
statistics from the largest ($J=1$) to the smallest ($J=1\mbox{,}536$). This
ranking procedure serves two purposes. First, it simplifies the
presentation of the results, since we can refer to each probe using its
ranking. Second, a pre-selection step using this ranking procedure is
used when evaluating the prediction performance in Section \ref{sec:data2}, which
is similar to what is often used in detecting differentially expressed
genes. We note that the use of Cox PH models is of no particular
importance, which simply provides a way to rank the probes; one can use
other models such as AFT models.

\subsection{Feature selection}\label{sec:data1}
Before building prediction scores, we conducted feature selection using
the following models: the Lasso-PL with $r=10$, Lasso-S$_K$, Lasso-L
and Lasso--Cox. In the Lasso-PL model \eqref{eq:plm.simp}, $X_i$ is
PSA, which is modeled using penalized splines, and $\bfZ$ includes the
binary clinical variable, gleason score, as well as the complete set or
a subset of 1,536 probes. Similarly, in the Lasso-S$_K$ model,
stratification is based on PSA.

\begin{table}
\caption{Feature selection for the prostate cancer study}\label
{tab:data-feature}
\begin{tabular*}{\tablewidth}{@{\extracolsep{\fill}}l@{\ \ }l@{}}
\hline
\textbf{Method} & \multicolumn{1}{c@{}}{\textbf{Selected probes}}\\
\hline
Lasso-PL & 1,\,2,\,4,\,12,\,16,\,31,\,38,\,46,\,63\\
Lasso-S$_2$& 1,\,4,\,8,\,12,\,16,\,31,\,46,\,63,\,382,\,906\\
Lasso-$S_4$& 1,\,4,\,12,\,16,\,29,\,31,\,36,\,38,\,46,\,56,\,70,\,78,\,310,\,382,\,390,\,591,\,1,500\\
Lasso-S$_8$& 1,\,4,\,8,\,9,\,16,\,18,\,31,\,36,\,37,\,38,\,46,\,56,\,57,\,70,\,78,\,178,\,237,\,271,\,310,\,855,\,1,500\\
Lasso-L & 1,\,2,\,4,\,8,\,9,\,16,\,31,\,46,\,63,\,70,\,136\\
Lasso--Cox & 2,\,4,\,8,\,11,\,14,\,16,\,22,\,31,\,46,\,52,\,63\\
\hline
\end{tabular*}
\end{table}

\begin{figure}[b]

\includegraphics{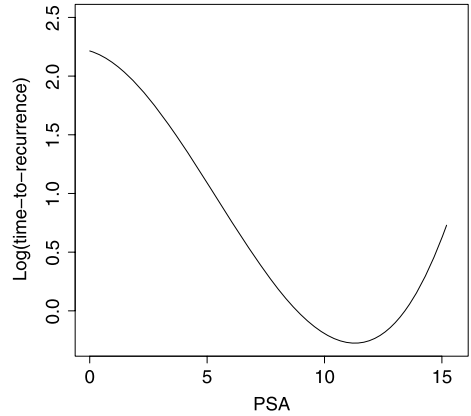}

\caption{Estimated nonlinear effect of PSA on the prostate cancer
recurrence after surgery~($\widehat\phi(X)$).}\label{fig:dat1}
\end{figure}

We first conducted an analysis using the complete set of 1,536 probes.
The results on feature selection are summarized in Table \ref
{tab:data-feature}. A linear effect of PSA was included in the Lasso-L
model and was estimated to be nonzero, which further justifies the
inclusion of PSA in other models; on the other hand, the total gleason
score is not selected by any of the methods. Figure \ref{fig:dat1}
shows the estimated effect of PSA using Lasso-PL; specifically, the
time to recurrence initially decreases as PSA increases and then starts
to increase slightly as PSA goes beyond 11. After further examination
of the data, we found that most patients had PSA values ranging from
0--15.2, but three had PSA values of 18.43, 26 and 32.10. More
importantly, all subjects with $\mathrm{PSA}>15.2$ had censored outcomes;
consequently, it is not appropriate to project the estimated $\phi(X)$
beyond 15.2. We also suspect that the increasing trend toward the right
tail is an artifact of the data and the effect of PSA instead levels
off when it is greater than 11, given that an increase in the time to
recurrence as PSA increases does not seem plausible clinically.

In terms of feature selection for the probe data, the Lasso-PL model
selects the least number of features, among which Probe 4, 16, 31 and
46 are selected by all six models, Probe 1 selected by five models,
Probe 63 selected by four models and Probe 2, 12 and 38 selected by
three models. In other words, all probes selected by Lasso-PL are
selected by at least half of all models, whereas other models select
some probes that are not shared by the rest of the models and are
likely to be noise. This agrees with the simulation results, that is,
in the presence of moderate to strong correlation among predictors, the
other models tend to select a larger number of noise features. In
addition, the difference between the Lasso-PL method and the Lasso-L
method is likely due to the nonlinear effect of PSA.

\subsection{Prediction performance}\label{sec:data2}
To internally evaluate the prediction performance, the data were
randomly split into a training sample (60\%) and a~validation sample
(40\%). Due to the high censoring rate, this step was stratified on the
censoring status to avoid extreme imbalance of censoring rates between
the training and validation samples. The models of interest were fit
using the training sample and were then used to construct the
predictive risk score for cancer recurrence, say, $\widehat\phi(X) +
\widehat\bfvtheta^{\mathrm{T}} \bfZ$ for Lasso-PL, for subjects in the
validation sample. Subsequently, the $c$ statistic was computed in the
validation sample. This procedure was repeated 1,000 times and the
average $c$ statistic is used for evaluating the prediction performance
of different models.

We compared the following model and data combinations: Lasso-PL with
$r=10$ using 1,536 probes and 2 clinical variables with PSA modeled
nonlinearly; Lasso-L and Lasso--Cox using 1,536 probes and 2 clinical
variables; Lasso-PL with $r=10$ using 2 clinical variables plus top 25
probes with PSA modeled nonlinearly, where the top 25 probes were
selected within each training sample; Lasso-L and Lasso--Cox using 2
clinical variables plus top~25 probes; partly linear AFT and Cox models
(PL-AFT and PL-Cox) using 2 clinical variables only with PSA modeled
nonlinearly through a penalized spline; linear AFT and Cox model (AFT
and Cox) using 2 clinical variables only. Note that we did not use
Lasso-S$_K$, since it does not estimate the nonlinear effect of PSA.

\begin{table}
\tablewidth=220pt
\caption{Prediction performance in the data analysis: mean $c$
statistic}\label{tab:datapred}
\begin{tabular*}{\tablewidth}{@{\extracolsep{\fill}}lcc@{}}
\\
\multicolumn{3}{@{}l@{}}{All 1,536 probes}\\
\hline
Lasso-PL & Lasso-L & Lasso--Cox \\
0.653 & 0.561  & 0.553 \\
\hline
\end{tabular*}\\[5pt]
\begin{tabular*}{\tablewidth}{@{\extracolsep{\fill}}lcc@{}}
\multicolumn{3}{@{}l@{}}{Top 25 probes}\\
\hline
Lasso-PL & Lasso-L & Lasso--Cox \\
0.653 & 0.567 & 0.572 \\
\hline
\end{tabular*}\\[5pt]
\begin{tabular*}{\tablewidth}{@{\extracolsep{\fill}}lccc@{}}
\multicolumn{4}{@{}l@{}}{Clinical variables only}\\
\hline
PL-AFT& AFT & PL-Cox & Cox \\
0.665 & 0.644 & 0.658 &0.644 \\
\hline
\end{tabular*}
\end{table}

Table \ref{tab:datapred} presents the mean $c$ statistic computed
using each model and data combination. Partly linear models have higher
average $c$ than linear models in all settings and for both AFT and Cox
models, indicating that the misspecified effect of PSA leads to worse
prediction performance. In all cases, AFT models have similar or higher
average $c$ compared to their corresponding Cox models. The average $c$
for Lasso-PL using all 1,536 probes is slightly less than PL-AFT using
only clinical variables, whereas Lasso-L and Lasso--Cox using all 1,536
probes have substantially lower $c$ than AFT and Cox using only
clinical variables. Furthermore, when a pre-selection step was included
to choose the top 25 probes first, we observe small improvement in $c$
for Lasso-L and Lasso--Cox and no improvement for Lasso-PL, which is
likely due to that the correctly modeled PSA effect plays the most
important role in prediction and the addition of gene expression data
does not seem to further improve prediction.

In summary, our analyses suggest that (1) the relationship between the
baseline PSA and prostate cancer recurrence is likely nonlinear, that
is, the time to recurrence decreases as PSA increases and it starts to
level off when PSA becomes greater than 11; (2) the correct
specification of this nonlinear effect improves performance in
prediction and feature selection; and (3) the addition of gene
expression data does not seem to further improve the prediction
performance. However, given that the sample size in this study is
small, our results need to be validated in a future study, preferably
with a larger sample size.

\section{Discussion}\label{sec:discussion}
We have investigated statistical approaches for prediction of clinical
end points that are subject to censoring. Our research shows that
correctly specifying nonlinear effects improves performance in both
prediction and feature selection for both low-dimensional and
high-dimensional data. While the proposed models can be used for
high-dimensional data, caution needs to be exercised in practice, since
the sample size is often small in real-life studies. This is especially
true when prediction is of primary interest and feature selection is
less of a concern. As the regularized methods achieve sparsity, they
shrink the coefficients of the important predictors. In finite samples,
such shrinkage becomes more pronounced as the noise level (i.e., the
number of noise predictors) increases; as a result, the prediction
performance deteriorates, which is reflected in our simulations and
data analysis.

We investigated two numerical methods for fitting proposed models. The
first algorithm is implemented through a $L_1$ regression, which is
slow for large data sets or when the number of predictors is large
relative to the sample size and fails when $d>n$. These limitations are
especially serious for censored data. For example, in our data example,
the first algorithm started to have convergence issues if $d>25$ probes
were used, in particular, when cross-validation was used or internal
validation was performed for evaluating prediction performance. The
second algorithm as described in Section \ref{sec:implementation} can
deal with high-dimensional data, and its solutions are fairly close to
those obtained using the first method when both are applicable.
Consequently, we recommend the use of the second algorithm in practice.

In this paper we focus on the performance for prediction as well as
feature selection in finite samples through extensive numerical
studies, and the theoretical properties of the proposed methods are
likely inherited from those of regularized linear AFT models and
penalized splines, which are beyond the scope of this article and are a
topic for future research. Nevertheless, our numerical results provide
empirical evidence to suggest that the proposed approach is likely to
enjoy the properties on feature selection that are possessed by
regularized estimation in linear AFT models [\citet{Cai09}] and in
stratified AFT models [\citet{Johnson09}].

Several metrics have been proposed for assessing the performance of
prediction models, and \citet{Steyerberg10} provide a nice review on
this subject; however, it is well known that censoring presents
additional challenges in developing such metrics [\citet{Begg00}; \citet{Gonen05}; \citet{Steyerberg10}]. In our simulations and data example, we
used the extended $c$ statistic to evaluate the prediction performance
in the presence of censored data; despite its ease of use, this metric
uses only concordant and disconcordant information and hence leads to
loss of information. Furthermore, while the existing metrics for
censored data are applicable for AFT models, no metric has been
proposed to take advantage of the unique feature of AFT models, namely,
they model the log-transformed outcome and can provide prediction on
the log-transformed scale, which is not trivial and is another topic
for our future research.

\section*{Acknowledgments}
We thank Editor Kafadar, an Associate Editor and two referees for their
helpful suggestions that greatly improved an earlier draft of this manuscript.


%

\printaddresses

\end{document}